\documentclass[sigconf,screen] {acmart}
\setcopyright{none}
\copyrightyear{2026}
\acmYear{2026}
\acmConference{CHI 2026 Workshop on Tools for Thought}{April 13--17, 2026}{Barcelona, Spain}
\setcopyright{none}
\usepackage[utf8]{inputenc}
\usepackage{graphicx}

\usepackage{subcaption}
\usepackage{enumitem}
\usepackage{array,booktabs}
\acmISBN{}
\acmDOI{}
\newcolumntype{L}[1]{>{\raggedright\arraybackslash}p{#1}}
\newcolumntype{Y}{>{\raggedright\arraybackslash}X}

\AtBeginDocument{}
\renewcommand\footnotetextcopyrightpermission[1]{}

\begin{document}

\title{Stop Writing for Me: Generative Refusal in AI Tools for Thought}

\author{Sora Kang}
\email{sorakang@snu.ac.kr}
\affiliation{%
  \institution{Seoul National University}
  \city{Seoul}
  \country{Republic of Korea}
}



\renewcommand{\shortauthors}{Sora Kang}

\begin{teaserfigure}
  \centering
  \includegraphics[width=\textwidth]{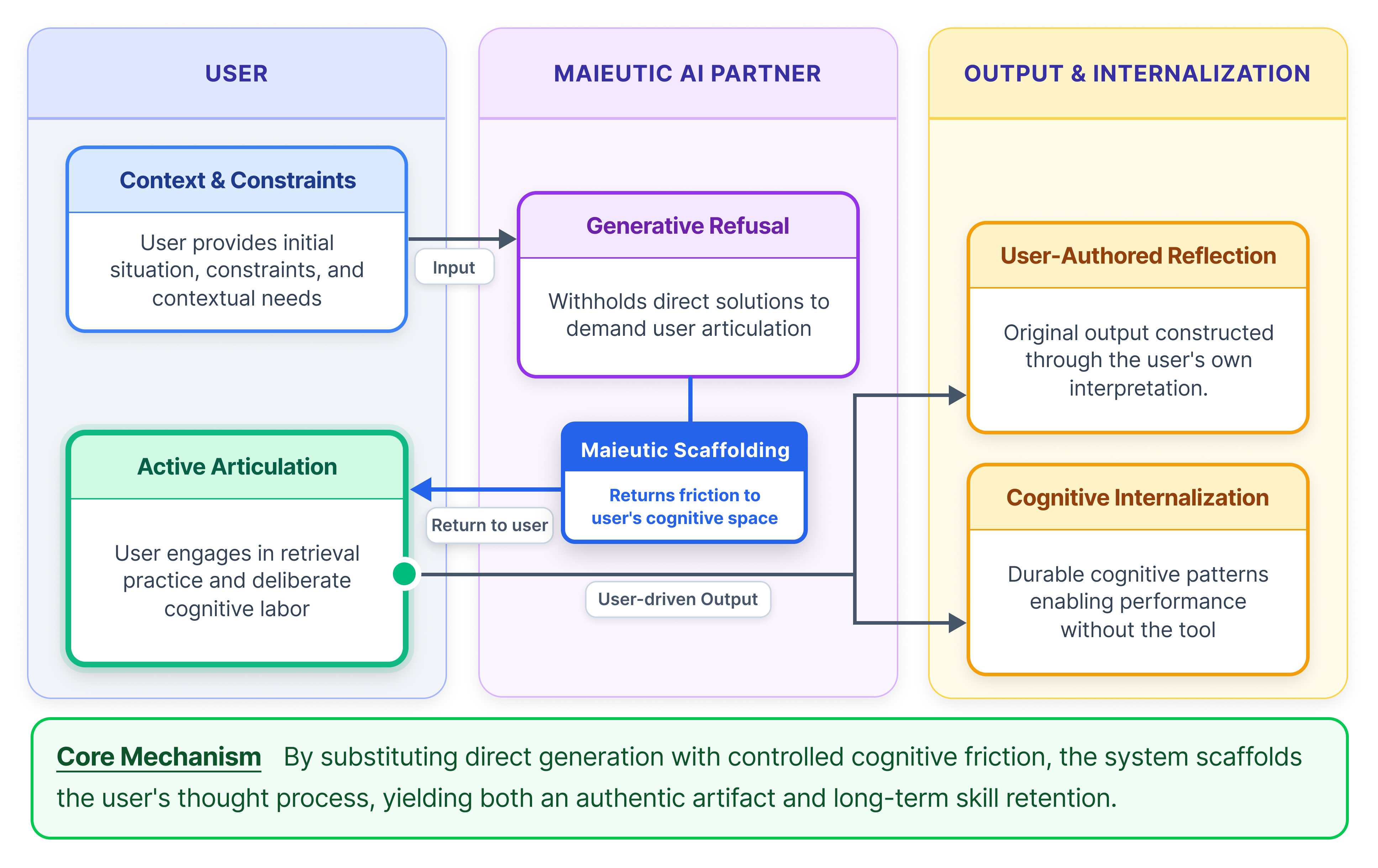}
  \caption{The Maieutic Interaction Framework. Instead of bypassing cognition, the system uses Generative Refusal to return cognitive friction to the user, shifting the interaction from delegation to active articulation.} 
  \Description{A node-link flowchart divided into three vertical swimlanes labeled 'User', 'Maieutic AI Partner', and 'Output & Internalization'. An arrow flows from 'Task Context & Constraints' in the User lane to 'Generative Refusal' in the AI lane. Instead of pointing directly to the output, the flow moves to 'Maieutic Scaffolding' and points backward into the User lane, connecting to a highlighted 'Active Articulation' node. From this node, two separate arrows point forward into the third lane, terminating at 'Authentic Artifact' and 'Cognitive Internalization'.}
  \label{fig:teaser}
\end{teaserfigure}

\begin{abstract}
In creative domains where the labor of articulation is central to the craft, how should we design Tools for Thought that enhance rather than bypass human cognition? Current GenAI paradigms often prioritize "cognitive offloading"—writing on behalf of users—risking the erosion of the constructive thought process essential to artistic training. In this position paper, we explore AI as a Maieutic Partner through "Generative Refusal"—strategically withholding text generation to demand user articulation. We discuss Actor’s Note, a journaling tool that generates context-aware questions instead of draft text. Our field study suggests that this constraint significantly reduced cognitive burden while fostering a residual effect of internalized questioning habits. We use these findings to discuss broader design implications for protecting human cognition against the tendency of generative efficiency.
\end{abstract}

\maketitle

\section{Introduction}
The concept of Tools for Thought (TfT) aims to extend human cognitive capabilities~\cite{bush1945as, clark1998extended, tankelevitch2025synthesis}. In the domain of actor training, character journaling—a pedagogical practice where actors write first-person narratives from their character's perspective to internalize inner motivations and emotional histories—serves as a foundational TfT for internalizing roles~\cite{bruder2012practical, cremin2006connecting, stanley2022actor, writinginrole, hancock1993character}. However, despite its established benefits, this practice is notoriously difficult to sustain~\cite{blix2015professional, norrthon2023knowledge, taylor2016actor}. Actors frequently abandon journaling not due to a lack of material, but because the high cognitive load required to initiate reflection after exhaustive rehearsals creates a significant barrier~\cite{bascomb2019performing, sorensen2024always}, often manifesting as the "blank page" problem~\cite{joyce2009blank, schultz1985writer}.

While Generative AI offers a potential solution to this sustainability gap~\cite{begus2024experimental, branch2021collaborative, dayo2023scriptwriting, dharaniya2023design}, current design paradigms prioritize efficiency through cognitive offloading—generating text on the user's behalf~\cite{chakrabarty2024creativity, coenen2021wordcraft}. Although this approach rapidly produces artifacts, it effectively bypasses the cognitive process of articulation. In domains like artistic training, where the mental effort of interpretation is central to learning~\cite{stanislavski2009actor, stanislavskij1986actor, hagen1991challenge}, such "write-for-me" features could deprive users of the opportunity for cognitive appropriation and internalization.

In this position paper, we argue that for GenAI to serve as a true TfT, it must shift from a "Co-author" role to that of a "Maieutic Partner" that facilitates inquiry.  This perspective resonates with recent HCI literature exploring "Socratic agents" and cognitive scaffolding \cite{metacognitive_cocreation_2025}. Recent studies demonstrate that intentionally employing a "refusal to generate" strategy---such as question-driven prewriting \cite{prewriting_asks_2026} or metacognitive prompting \cite{socratic_chatbot_2024, socratic_highered_2025}---effectively promotes critical thinking over cognitive offloading. Furthermore, Socratic AI has shown promise in mitigating evaluation anxiety in high-stakes training environments \cite{nursing_socratic_2024}. However, existing work has primarily applied this paradigm to objective academic tasks or structured prewriting. 

To demonstrate how this Socratic approach can be extended into the arts—specifically, the subjective, embodied, and emotionally demanding domain of artistic training—we discuss Actor’s Note, a work to be presented at CHI 2026~\cite{kang2026actors}. This system implements a constraint-based design that strictly refuses to generate text, instead providing context-aware questions tailored to the actor’s rehearsal stage. By employing "good friction"~\cite{fleck2010reflecting}, the system lowers the cognitive burden of initiation while augmenting the depth of thought. Based on empirical findings from this implementation, we aim to discuss broader design implications for future TfTs, each accompanied by a workshop discussion question: (1) employing Generative Refusal to demand user articulation rather than filling the gap with synthetic text; (2) implementing Adaptive Scaffolding that adjusts friction based on the user's workflow phase; (3) leveraging AI as a Non-judgmental Space to mitigate social evaluation threats; and (4) adopting Internalization—performance without the tool—as a primary success metric.

\section{Case Study: Actor's Note}
We developed Actor's Note to operationalize the vision of Maieutic AI and validated it through a rigorous field deployment.

\subsection{System Overview}
Actor's Note is a web-based journaling tool powered by a Large Language Model (GPT-4o). Its core design principle is the refusal to generate text on behalf of the user. Instead, the system analyzes the user's uploaded script, role, and performance date to generate three context-specific questions daily. Crucially, the system utilizes the actor's current rehearsal stage as a prompt parameter—focusing, for example, on backstory details during table work versus emotional continuity during run-throughs. These questions are framed in the second person to elicit first-person reflection, ensuring the cognitive labor of interpretation remains with the user.

\subsection{Study Overview}
To evaluate the system, we conducted a 14-day in-the-wild study with 29 professional and student actors using a randomized crossover design. Following a baseline period to establish journaling habits, participants were counterbalanced across two conditions: an AI-assisted phase using Actor's Note and an unassisted freewriting phase. We collected interaction logs to analyze writing latency and linguistic diversity, alongside daily surveys measuring cognitive burden, intrinsic motivation, and acting confidence . Semi-structured interviews were conducted post-study to examine qualitative shifts in the actors' reflective processes.

\subsection{Key Findings}
The immediate effect of the system was a simultaneous reduction in entry barriers and an improvement in the quality of reflection. Despite maintaining the friction of writing, results showed a significant decrease in Cognitive Burden ($\eta_p^2=.381, p<.001$) and an increase in Intrinsic Motivation ($\eta_p^2=.201, p=.017$). Acting Confidence also increased significantly ($\eta_p^2=.352, p<.001$), suggesting that the AI-generated questions concretized character interpretation. Beyond subjective metrics, log analysis confirmed a deepening of inquiry; entries written with AI assistance exhibited significantly higher Lexical Diversity ($q=.025$) and increased usage of both negative ($q=.001$) and positive ($q=.025$) emotion words compared to unassisted entries, indicating that the questions prompted actors to articulate a broader and more complex range of emotions.

Crucially, the tool's impact varied by timing and fostered lasting cognitive habits. For the group introducing AI early, the tool functioned as a momentum starter that rapidly reduced cognitive burden, whereas for the group introducing AI later, it functioned as a deepener, leading to a significant increase in Narrative Transportation ($p=.0128$). This suggests friction plays a different role depending on the user's workflow stage. Furthermore, post-study data indicates that the questioning framework was internalized; after AI assistance was removed, participants reported a high tendency to recall the AI's questioning style or self-generate similar questions ($M=4.87$ on a 7-point scale), demonstrating that the intervention fostered a lasting cognitive habit rather than a temporary dependency.

\section{Discussion}
The deployment of Actor's Note demonstrates that constraining generative capabilities can, counter-intuitively, enhance user engagement and cognitive depth. We offer four implications for designing GenAI tools that aim to support, rather than replace, human thought processes.

\subsection{Generative Refusal}
The prevailing design logic in Generative AI prioritizes the reduction of friction, aiming to produce the final artifact with minimal user effort. However, our findings suggest that in educational and training contexts, this efficiency is detrimental to learning. By implementing a "Refusal to Generate," Actor's Note forced users to engage in retrieval practice and articulation—cognitive processes essential for memory consolidation and understanding. We argue that "Generative Refusal" should be treated not as a lack of functionality, but as a core mechanic for Tools for Thought. Designers should utilize GenAI to create a structured gap that demands user input, rather than filling that gap with synthetic text. This shifts the interaction model from "delegation" (AI does it for me) to "activation" (AI prompts me to do it).

This concept of Generative Refusal aligns closely with Bjork's notion of ``desirable difficulties'' \cite{bjork1994memory}, which posits that introducing controlled cognitive friction into the learning process may slow down immediate performance but significantly enhances long-term retention and the transfer of skills. By intentionally refusing to bypass the ``blank page'' for the user, Actor's Note introduces this desirable difficulty. Furthermore, the AI-generated questions serve as cognitive scaffolding within the user's Zone of Proximal Development (ZPD) \cite{vygotsky1978mind}, guiding the actors to construct their own interpretations rather than passively consuming synthetic text.

\textbf{Workshop Question:} \textit{Where lies the "optimal friction point" that balances the immediate desire for productivity with the long-term necessity of cognitive depth, and how can we design interfaces that help users negotiate this trade-off?}

\subsection{Adaptive Scaffolding}
Our study revealed that the user's cognitive needs shift significantly as their work progresses. In the early stages of rehearsal, users struggled with inertia and the "blank page," requiring low-friction prompts to initiate writing. In later stages, however, users struggled with fixation on established interpretations, requiring high-friction, challenging questions to deepen their understanding. A static tool cannot address both needs effectively. Future TfT systems must be context-aware, adjusting their scaffolding strategy dynamically based on the user's workflow phase. Detecting this phase presents a key design challenge, and we envision several potential approaches to explore. For example, a system might employ (1) \textbf{System-driven inference}, estimating progress based on script metadata or temporal proximity to the performance date; alternatively, it could rely on (2) \textbf{User-driven selection}, allowing actors to self-report their current rehearsal stage (e.g., `table work' vs. `run-through') prior to journaling; or even (3) \textbf{Leader-driven synchronization}, where a director manually updates the global phase for the entire cast. Depending on the detected phase, the system could then act as a pacemaker in the beginning to build momentum, and as a challenger or devil's advocate in later stages to prevent cognitive stagnation.

\textbf{Workshop Question:} \textit{How might this "Maieutic" form of questioning be operationalized in other creative domains like coding or academic writing—where might "refusal" look like a syntax error, and where might it look like a helpful prompt?}

\subsection{Mitigating Social Evaluation Threat}
Deep reflection requires vulnerability, yet human feedback loops often introduce a social evaluation threat that leads to self-censorship. Our participants explicitly noted that they shared raw, unpolished thoughts with the AI that they would hide from human directors or peers to maintain their professional image. This indicates that AI's lack of social agency is a functional asset. Designers should explicitly frame AI-based reflection tools as "non-judgmental spaces" where the cost of error is zero. By removing the fear of social judgment, AI can facilitate a level of honest introspection and experimentation that is often inhibited in human-to-human training environments. However, we must also acknowledge an inherent tension here: the very absence of social stakes that makes AI a safe space might also reduce the extrinsic motivation, adrenaline, and performative energy typically derived from a human audience or director. Therefore, AI-based reflection tools should be positioned not as replacements for the rehearsal floor, but as preparatory sandboxes. Future research should explore how the private introspection facilitated by AI can be effectively bridged with the necessary, high-stakes environment of live performance, ensuring these tools complement rather than isolate actors from human feedback loops.

\textbf{Workshop Question:} \textit{Could the "Character Journaling" method—adopting a persona to bypass self-censorship—be generalized to other high-stakes professional contexts (e.g., executive coaching, therapy), and does the non-human nature of AI make it the ideal mediator for this specific technique?}

\subsection{Internalization as the Success Metric}

The current industry standard for evaluating GenAI focuses on performance with the tool—measuring how much faster or better a user can produce an output while using the AI. However, this metric fails to distinguish between genuine cognitive augmentation and dependency. Our study measured "performance without the tool" (residual effects) and found that users internalized the questioning patterns, applying them independently after the intervention. We propose that the HCI community must adopt "Internalization" as a primary success metric for TfTs. If a user cannot perform the task better after the tool is removed, the system has merely served as a prosthetic for production, not a scaffold for thought.

\textbf{Workshop Question:} \textit{If "Post-Tool Performance" is the true measure of a TfT, what specific longitudinal metrics or "unassisted" evaluation protocols should the HCI community adopt to standardize the measurement of cognitive internalization?} 

\section{Conclusion}
We presented Actor's Note to demonstrate that "Generative Refusal" is a viable design strategy for protecting cognitive engagement. To serve as a potential blueprint for future Tools for Thought, we propose four design shifts: (1) treating refusal as an active feature to demand user articulation, (2) implementing adaptive scaffolding that shifts from support to challenge based on workflow, (3) leveraging AI as a non-judgmental space to mitigate social evaluation threat, and (4) adopting internalization (post-tool performance) as a primary success metric. We hope the workshop discussion will encourage the HCI community to move beyond prosthetic AI toward systems that support independent human capabilities.

\bibliographystyle{ACM-Reference-Format}
\bibliography{chi26-workshop.bib}

\end{document}